\documentclass{elsart}

\usepackage{graphicx}

\usepackage{amssymb}

\begin{document}

\begin{frontmatter}



\title{On the stability of optical lattices}


\author[ON]{G. Di Domenico\corauthref{cor1}},
\ead{Gianni.Didomenico@ne.ch}
\author[ON]{N. Castagna},
\author[ON]{M.D. Plimmer},
\author[ON]{P. Thomann},
\author[LLF1,LLF2]{A.V. Taichenachev},
\author[LLF1,LLF2]{V.I. Yudin}

\address[ON]{Observatoire cantonal, rue de l'Observatoire 58, 2000 Neuch\^{a}tel, Switzerland}
\address[LLF1]{Novosibirsk State University, Pirogova 2, Novosibirsk 630090, Russia}
\address[LLF2]{Institute of Laser Physics SB RAS, Lavrent'eva 13/3, Novosibirsk 630090, Russia}

\corauth[cor1]{Corresponding author.}

\begin{abstract}
In this article, we present an analysis of the stability of
optical lattices. Starting with the study of an unstable optical
lattice, we establish a necessary and sufficient condition for
intrinsic phase stability, and discuss two practical solutions to
fulfill this condition, namely minimal and folded optical
lattices. We then present a particular example of two-dimensional
folded optical lattice which has the advantage of being symmetric,
power recycling and having a convenient geometry. We have used
this lattice for laser collimation of a continuous cesium beam in
a fountain geometry.
\end{abstract}

\end{frontmatter}

\section{Introduction}
\label{sec:introduction}

It is well known that the energy levels of an atom interacting
with an electromagnetic field undergo an a.c. Stark or light shift
\cite{Barrat61,Happer67}, proportional to the local light
intensity. In a standing wave, the ground state light shift gives
rise to a periodic potential, called an optical lattice, which can
be used to trap the atoms \cite{Jessen96,Guidoni99,Rolston98}. The
first experiments demonstrating the trapping  of atoms in the
potential wells of an optical lattice and their localisation
therein were performed by the groups of Grynberg \cite{Lounis92a}
and Jessen \cite{Jessen92}. Since then, studies of laser cooling,
quantum state preparation, and Bose-Einstein condensates in
optical lattices have intensified.

In this article, we present a general analysis of the stability of
optical lattices. After a short presentation of the instability
problem, we establish a necessary and sufficient condition for
intrinsic phase stability. Then we discuss two practical solutions
to guarantee intrinsic phase stability: the first of these,
minimal optical lattices, is the solution proposed by Grynberg
{\em et al.} in 1993 \cite{Grynberg93}, while the second, folded
optical lattices, was suggested by Rauschenbeutel {\em et al.} a
few years later \cite{Rauschenbeutel98}. Finally, we discuss both
approaches and describe a laser cooling experiment in which we
have used a two-dimensional (2D) folded optical lattice for the
collimation of a continuous atomic beam.

In the following, we consider an optical lattice resulting from
the superposition of $l$ laser beams. The total electric field is
given by $\mathbf{E}_L(\mathbf{r},t)
=\mathrm{Re}\left[\mathbf{E}_L(\mathbf{r})\exp({-i\omega_L t
})\right]$ where
\begin{equation}
  \label{equ:Lattice_laser_beams}
  \mathbf{E}_L(\mathbf{r})=
  \sum_{j=1}^{l}E_j\,\vec{\varepsilon}_j\,\exp\left[{i(\mathbf{k}_j\cdot\mathbf{r}+\phi_j)}\right]
  \;,
\end{equation}
$\phi_j$ being the phase of the $j$th laser beam. As explained in
Refs.~\cite{Jessen96} and \cite{Deutsch98}, the optical shift
operator for atoms in the ground state is given by
\begin{equation}
  \label{equ:Lattice_LS_operator}
  \hat{U}(\mathbf{r})=
  -\mathbf{E}^{\ast}_L(\mathbf{r})\cdot\mathbf{\hat{\alpha}}\cdot\mathbf{E}_L(\mathbf{r})
\end{equation}
where $\mathbf{\hat{\alpha}}$ is the atomic polarisability tensor
operator given by
$\mathbf{\hat{\alpha}}=-\sum_e\mathbf{\hat{d}}_{ge}\mathbf{\hat{d}}_{eg}/\hbar\Delta_{ge}$
where $\Delta_{ge}$ is the laser detuning with respect to the
atomic transition $|g\rangle\rightarrow|e\rangle$ and
$\mathbf{\hat{d}}_{eg}$ is the electric dipole operator between
these levels. By inserting Eq.~(\ref{equ:Lattice_laser_beams})
into Eq.~(\ref{equ:Lattice_LS_operator}) we get
\begin{eqnarray}
  \label{equ:Lattice_light_shift_phases}
  \hat{U}(\mathbf{r})&=&
  -\sum_{i,j}\,(\vec{\varepsilon}_{i}^{\,\ast}\cdot\mathbf{\hat{\alpha}}\cdot\vec{\varepsilon}_{j})
  E^{\ast}_i E_j\,
  \nonumber\\
  & &\times
  \exp\left[{i(\phi_j-\phi_i)}\right]
  \exp\left[{i(\mathbf{k}_j-\mathbf{k}_i)\cdot\mathbf{r}}\right],
\end{eqnarray}
which is our starting point for the stability analysis.

Although in the following stability analysis we concentrate on the
optical shift operator (\ref{equ:Lattice_LS_operator}), all other
field dependent operators of the problem (e.g. the optical pumping
rate operator) are also determined by quadratic combinations of
the type (\ref{equ:Lattice_light_shift_phases}). Therefore, the
stability conditions will be the same for these operators too.

\section{Problem of stability}
\label{sec:problem}

From Eq.~(\ref{equ:Lattice_light_shift_phases}), it is clear that
the relative phases $\phi_j-\phi_i$ play a critical role. Any
variation in one of these phases arising, for example, from the
vibration of a mirror, may manifest itself as a dramatic change in
the optical potential.

\begin{figure}[t!]
  \centering
  \includegraphics[width=9cm]{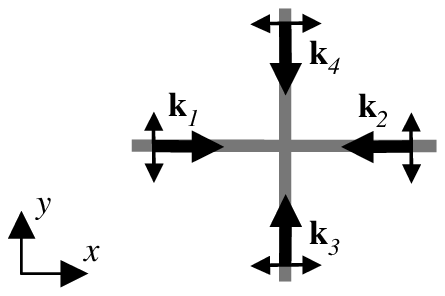}
  \caption{Example of an unstable optical lattice of dimension 2.}
  \label{fig:Lattice_UnstableLattice}
\end{figure}

To illustrate this problem, let us consider the 2D optical lattice
shown in Fig.~\ref{fig:Lattice_UnstableLattice}, where four laser
beams intersect in a common plane, all linearly polarised within
that plane.

\begin{figure}[t!]
  \centering
  \includegraphics[width=9cm]{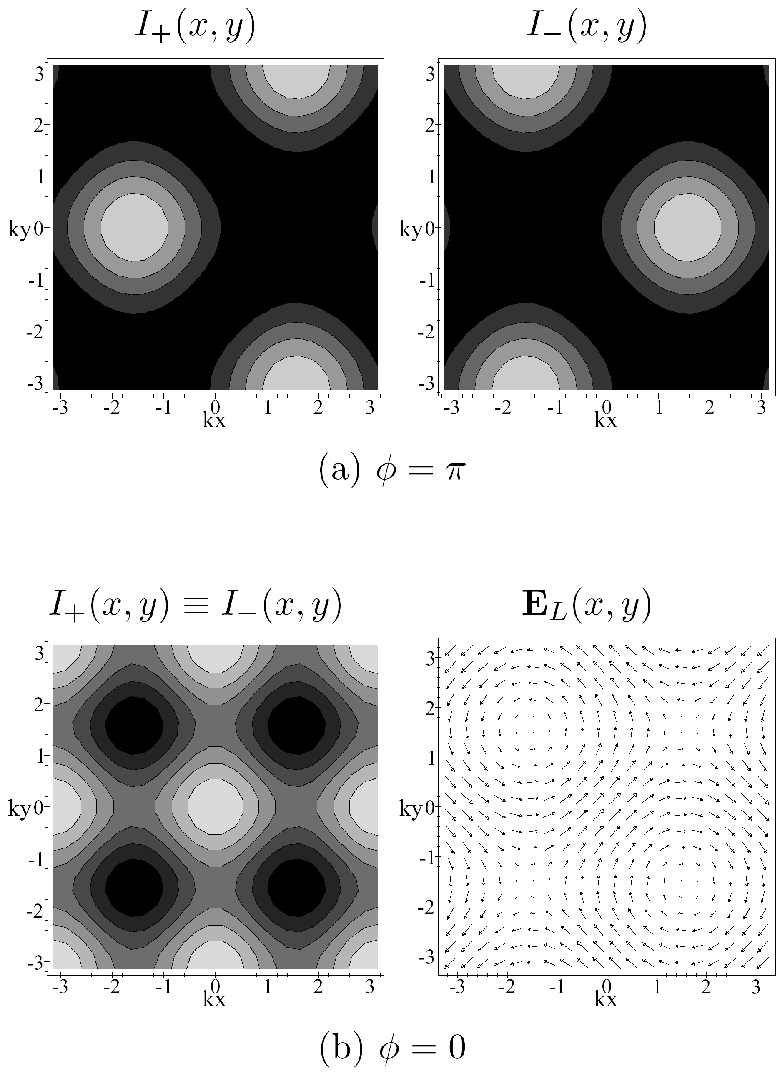}
  \caption{Representation of the polarisation gradients
  for the 2D optical lattice shown in Fig.~\ref{fig:Lattice_UnstableLattice}.
  (a) Case where the phase of the first laser beam is $\phi=\pi$.
  We have plotted the intensity of circular polarisation
  components $I_{\pm}(x,y)$ from Eq.~(\ref{equ:Ipm_pi}).
  In this case, sites of pure $\sigma_+$ and $\sigma_-$
  polarisation are in alternance every half-wavelength.
  (b) Case where the phase of the first laser beam is $\phi=0$.
  On the left we have plotted $I_{\pm}(x,y)$ from Eq.~(\ref{equ:Ipm_zero}).
  In this case $I_{+}(x,y)\equiv I_{-}(x,y)$, therefore the polarisation is linear
  everywhere. However, there is still a strong polarisation gradient,
  as can be seen on the right graph where we have plotted the
  polarisation vector field $\mathbf{E}_L(x,y)$ in the lattice plane.
  }
  \label{fig:Lattice_UnstableLatticeIGraph}
\end{figure}

The total electric field is given by the superposition of the
electric fields of the four laser beams (with identical amplitude
$E$):
\begin{eqnarray}
  \mathbf{E}_L(x,y)&=&
  E \,\left[\,
    \vec{\varepsilon}_y \exp({ikx+i\phi})
  + \vec{\varepsilon}_y \exp({-ikx})\right.
  \nonumber\\*
  & & \left.
  \;+\,\vec{\varepsilon}_x \exp({iky})
  + \vec{\varepsilon}_x \exp({-iky})
  \,\right]
  \label{equ:Lattice_unstable_lattice_E_tot}
\end{eqnarray}
where we have introduced the variable $\phi$ to represent a change
in the phase of the first laser beam. Grouping terms with
identical polarisation, we obtain:
\begin{equation} \label{equ:Lattice_unstable_lattice_E_tot_bis}
  \mathbf{E}_L(x,y) = 2E
  \left[\,
  \vec{\varepsilon}_x \cos(ky)
  +\vec{\varepsilon}_y \exp({i\phi/2})\cos\left(kx+\phi/2\right)
  \right].
\end{equation}
This expression shows that the total polarisation is critically
dependent on the angle $\phi$. Indeed, if we calculate the
intensity of circular polarisation components
$I_{\pm}(x,y)=|\,\vec{\varepsilon}_{\pm}^{\,\ast}\cdot\mathbf{E}_L(x,y)|^2$
where
$\vec{\varepsilon}_{\pm}=\frac{1}{\sqrt{2}}(\vec{\varepsilon}_x\pm
i\vec{\varepsilon}_y)$ we obtain, for $\phi=\pi$:
\begin{equation}
  \label{equ:Ipm_pi}
  I_{\pm}(x,y)=
  2E^2\left[\cos(ky)\mp\sin(kx)\right]^2
\end{equation}
and for $\phi=0$:
\begin{equation}
  \label{equ:Ipm_zero}
  I_{\pm}(x,y)=
  2E^2\left[\cos^2(ky)+\cos^2(kx)\right]\,.
\end{equation}
We have plotted these circular polarisation components in
Fig.~\ref{fig:Lattice_UnstableLatticeIGraph}. These two situations
are very different.

For $\phi=\pi$, we observe alternating sites of pure $\sigma_+$
and $\sigma_-$ polarisation with a period equal to the wavelength.
This means the light shifts of $m_F\!=\!\pm F$ Zeeman sub-levels
are periodic in space with opposite phases, and the same is true
for the optical pumping process which always populates the ground
state $m_F$ sub-level of lowest energy. As a consequence, Sisyphus
cooling \cite{dalibard89} can take place if the laser beams are
tuned correctly.

On the contrary, for $\phi=0$ we have $I_{+}(x,y)\equiv
I_{-}(x,y)$, both having a period equal to a half-wavelength. The
polarisation is thus linear everywhere, which precludes Sisyphus
cooling. However, there is still a strong polarisation gradient,
as can be seen in the right graph of
Fig.~\ref{fig:Lattice_UnstableLatticeIGraph}(b) where we have
plotted the polarisation vector as a function of position in the
optical lattice plane. Thus, it is probable that another
sub-Doppler cooling mechanism takes place in this situation, for
example a mechanism similar to that operating in
$\sigma^+$-$\sigma^-$ molasses \cite{dalibard89}.

We conclude that the instability of the optical lattice leads to
dramatic changes of the polarisation gradients. This is a crucial
problem when one works with cooling mechanisms involving the
spatial dependence of light polarisation. One solution to this
problem is to stabilize mechanically the phase difference between
laser beams, as was first implemented by Hemmerich \textit{et al.}
\cite{Hemmerich93}. However, other approaches free of this
mechanical constraint were proposed by the groups of Grynberg
\cite{Grynberg93} and Meschede \cite{Rauschenbeutel98}. We shall
discuss both of these approaches in sections \ref{sec:minimal} and
\ref{sec:folded}, but first we start by defining intrinsic phase
stability and by establishing a necessary and sufficient condition
for it.

\section{Necessary and sufficient condition for intrinsic phase stability}
\label{sec:iif}

Let us consider an optical lattice composed of $l$ laser beams,
and let us suppose that the phases of these laser beams change
suddenly as follows
\begin{equation}
  \phi_j\rightarrow\phi_j+\Delta\phi_j\,.
\end{equation}
From equation (\ref{equ:Lattice_light_shift_phases}), we see that
it is possible to compensate the effect of this change by a
translation $\mathbf{r}\rightarrow\mathbf{r}-\mathbf{\Delta r}$,
provided there exists a vector $\mathbf{\Delta r}$ and an
arbitrary phase $\phi_o$ (which has no effect on
$\hat{U}(\mathbf{r})$) such that
\begin{equation}\label{equ:Lattice_syst_lin_phases1}
  \mathbf{k}_j\cdot\mathbf{\Delta r} = \Delta\phi_j+\phi_o,\,\forall j=1,\ldots ,l \,.
\end{equation}
This brings us to the following definition, namely, we say that an
optical lattice is {\em intrinsically phase stable } if any change
in the phases of the laser beams can be compensated by a formal
translation, as explained above. In practice, this means that
every change in the phases of the laser beams manifests itself as
a physical translation of the optical lattice in space. Such
translations do not disturb the optical cooling mechanisms as long
as the atoms' internal variables evolve much more rapidly than the
optical potential, which is usually the case.

Let us state now an important result. We denote by $\mathcal{K}$
the matrix composed of the lines $(\mathbf{k}_j-\mathbf{k}_1)^T$
for $j=2,\ldots,l$, and $\mathbf{\Phi}$ the vector composed of the
elements $\Delta\phi_j-\Delta\phi_1$ for $j=2,\ldots,l$. We recall
that the rank of a matrix is equal to the dimension of the vector
space generated by its columns or by its rows. Then, the optical
lattice resulting from the superposition of the $l$ laser beams is
intrinsically phase stable if and only if
\begin{equation}
  \label{equ:rank_stability_condition}
  \mathrm{rank}(\mathcal{K}) =
  \mathrm{rank}\left(\mathcal{K}|\mathbf{\Phi}\right),\,\forall \mathbf{\Phi}
\end{equation}
where $\left(\mathcal{K}|\mathbf{\Phi}\right)$ is the matrix
obtained from $\mathcal{K}$ by adding a column composed of the
elements of $\mathbf{\Phi}$.

The proof of this result is as follows. It is clear that the
optical lattice is intrinsically phase stable if and only if the
linear system (\ref{equ:Lattice_syst_lin_phases1}) always admits a
solution, for any choice of the phase variations $\Delta\phi_j$.
By subtraction of the first equation, we obtain the following
linear system
\begin{equation}\label{equ:Lattice_syst_lin_phases2}
  \left\{(\mathbf{k}_j-\mathbf{k}_1)\cdot\mathbf{\Delta r} = \Delta\phi_j-\Delta\phi_1\,|\,j=2,\ldots,l\right\}
\end{equation}
which is equivalent to the system
(\ref{equ:Lattice_syst_lin_phases1}). The system
(\ref{equ:Lattice_syst_lin_phases2}) can be written in matrix form
as $\mathcal{K}\mathbf{\Delta r}=\mathbf{\Phi}$. For this equation
to have at least one solution, it is both necessary and sufficient
that $\mathrm{rank}(\mathcal{K})$ be equal to
$\mathrm{rank}\left(\mathcal{K}|\mathbf{\Phi}\right)$ (this is the
Rouch\'{e}-Capelli theorem for the existence of the solution of a
linear system). Here ends the proof.

\section{Minimal optical lattices}
\label{sec:minimal}

Let us consider an optical lattice composed of $l$ laser beams,
and let us denote $d$ its spatial dimension. We are going to
demonstrate that $d=l-1$ is a sufficient condition to guarantee
the intrinsic phase stability of the optical lattice. Let us start
by a demonstration of the following properties:
\begin{enumerate}
  \item $d=\mathrm{rank}(\mathcal{K})$;
  \item $\mathrm{rank}(\mathcal{K})\leq\mathrm{rank}\left(\mathcal{K}|\mathbf{\Phi}\right)$;
  \item $\mathrm{rank}\left(\mathcal{K}|\mathbf{\Phi}\right)\leq l-1$.
\end{enumerate}

As can be seen from Eq.~(\ref{equ:Lattice_light_shift_phases}),
the optical lattice is generated by the vectors
$(\mathbf{k}_j-\mathbf{k}_i)$, thus its spatial dimension $d$ is
equal to the dimension of the vector space generated by the
vectors $(\mathbf{k}_j-\mathbf{k}_i)$. But this vector space is
also generated by the vectors $(\mathbf{k}_j-\mathbf{k}_1)$ which
compose the matrix $\mathcal{K}$. Therefore, by definition of the
rank, $d$ is equal to $\mathrm{rank}(\mathcal{K})$, which proves
the first property. The second property comes from the trivial
assertion that adding a column to a matrix cannot decrease the
rank. Finally, the rank of a matrix cannot exceed the number of
lines, and this proves the last inequality.

Writing these three properties side by side, we get:
\begin{equation}\label{equ:Lattice_inegalite_rank}
  d=\mathrm{rank}(\mathcal{K})
  \leq
  \mathrm{rank}\left(\mathcal{K}|\mathbf{\Phi}\right)
  \leq
  l-1\,.
\end{equation}
From this expression, it is obvious that by imposing the condition
$d=l-1$ we guarantee that
$\mathrm{rank}(\mathcal{K})=\mathrm{rank}\left(\mathcal{K}|\mathbf{\Phi}\right)$
and therefore that the optical lattice is intrinsically phase
stable. This is the solution proposed by Grynberg \textit{et al.}
in 1993 to build stable optical lattices \cite{Grynberg93}. Note
that Eq.~(\ref{equ:Lattice_inegalite_rank}) implies $l\geq d+1$.
Therefore, $l=d+1$ is the \textit{minimum} number of laser beams
needed to create an optical lattice of dimension $d$, hence the
term \textit{minimal optical lattice}.

At this point, it is important to note that the condition $d=l-1$
is sufficient, but not necessary, to have
$\mathrm{rank}(\mathcal{K})=\mathrm{rank}\left(\mathcal{K}|\mathbf{\Phi}\right)$.
There is another method to obtain an optical lattice which is
intrinsically phase stable. It is described in the next section.

\section{Folded optical lattices}
\label{sec:folded}

\begin{figure}[t!]
  \centering
  \includegraphics[width=9cm]{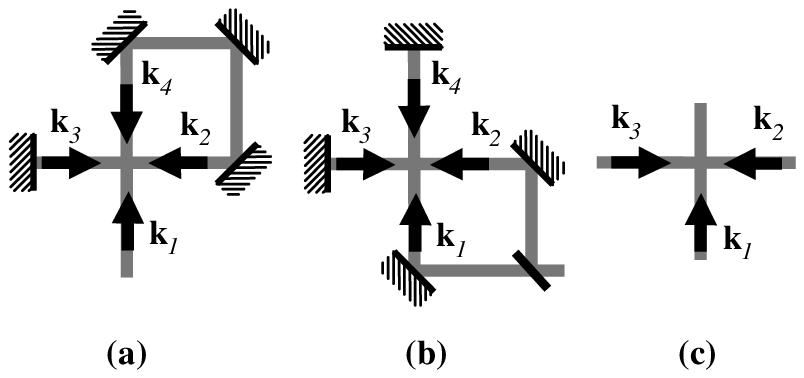}
  \caption{Comparison of optical lattices of dimension $d=2$: (a) folded optical lattice; (b) unstable optical lattice; (c) minimal optical lattice (with $d=l-1$).}
  \label{fig:Lattice_abc_lattice_schemes}
\end{figure}

Let us consider the 2D optical lattice geometry presented in
Fig.~\ref{fig:Lattice_abc_lattice_schemes}(a). This optical
lattice is intrinsically phase stable, even though it does not
satisfy the condition $d=l-1$. To explain this point, we start by
considering the optical lattice of
Fig.~\ref{fig:Lattice_abc_lattice_schemes}(c) which is composed of
the first three laser beams $\mathbf{k}_1$, $\mathbf{k}_2$ and
$\mathbf{k}_3$. This optical lattice is intrinsically phase stable
since it satisfies the condition $d=l-1$. Therefore, we have
\begin{equation}\label{equ:Lattice_stab_3_lasers}
  \mathrm{rank}\!
  \left(
  \begin{array}{c}
    \mathbf{k}_2^T\!-\!\mathbf{k}_1^T \\
    \mathbf{k}_3^T\!-\!\mathbf{k}_1^T
  \end{array}
  \right)
  =
  \mathrm{rank}\!
  \left(
  \begin{array}{c|c}
    \mathbf{k}_2^T\!-\!\mathbf{k}_1^T & \Delta\phi_2\!-\!\Delta\phi_1 \\
    \mathbf{k}_3^T\!-\!\mathbf{k}_1^T & \Delta\phi_3\!-\!\Delta\phi_1
  \end{array}
  \right)
  =2
\;.
\end{equation}
Let us now reconsider the optical lattice of
Fig.~\ref{fig:Lattice_abc_lattice_schemes}(a). Since the
retro-reflected beam follows the same path as the incident beam,
the phases satisfy the relation $\phi_4-\phi_3=\phi_2-\phi_1$.
Therefore, the differences $\phi_j-\phi_1$ are linked via
\begin{equation}\label{equ:Lattice_phase_relation}
  (\phi_4-\phi_1)=(\phi_3-\phi_1)+(\phi_2-\phi_1)
  \;.
\end{equation}
On the other hand, the differences $\mathbf{k}_j-\mathbf{k}_1$ are
related by
\begin{equation}\label{equ:Lattice_kj_relation}
  (\mathbf{k}_4-\mathbf{k}_1)=(\mathbf{k}_3-\mathbf{k}_1)+(\mathbf{k}_2-\mathbf{k}_1)
  \;.
\end{equation}
Since the linear combinations (\ref{equ:Lattice_phase_relation})
and (\ref{equ:Lattice_kj_relation}) are identical, we have
\begin{eqnarray}
  &&\mathrm{rank}\!
  \left(
  \begin{array}{c|c}
    \mathbf{k}_2^T\!-\!\mathbf{k}_1^T & \Delta\phi_2\!-\!\Delta\phi_1 \\
    \mathbf{k}_3^T\!-\!\mathbf{k}_1^T & \Delta\phi_3\!-\!\Delta\phi_1 \\
    \mathbf{k}_4^T\!-\!\mathbf{k}_1^T & \Delta\phi_4\!-\!\Delta\phi_1
  \end{array}
  \right)\nonumber \\
  =&&\mathrm{rank}\!
  \left(
  \begin{array}{c|c}
    \mathbf{k}_2^T\!-\!\mathbf{k}_1^T & \Delta\phi_2\!-\!\Delta\phi_1 \\
    \mathbf{k}_3^T\!-\!\mathbf{k}_1^T & \Delta\phi_3\!-\!\Delta\phi_1
  \end{array}
  \right)
\end{eqnarray}
and thus $\mathrm{rank}\left(\mathcal{K}|\mathbf{\Phi}\right)=2$.
Now, using Eq.~(\ref{equ:Lattice_inegalite_rank}) with $d=2$, we
can conclude that
$\mathrm{rank}(\mathcal{K})=\mathrm{rank}\left(\mathcal{K}|\mathbf{\Phi}\right)$
is always satisfied, and therefore the optical lattice of
Fig.~\ref{fig:Lattice_abc_lattice_schemes}(a) is intrinsically
phase stable.

The idea of using this type of intrinsically phase stable
configuration was initially put forward by Rauschenbeutel {\em et
al.} in a slightly different form \cite{Rauschenbeutel98}. Their
explanation for stability is more intuitive and consists in
observing that the optical lattice of
Fig.~\ref{fig:Lattice_abc_lattice_schemes}(a) is created by
folding a 1D lattice such that it intersects with itself. Since 1D
lattices are instrinsically stable, as discussed above,  folded
ones must be too.

One can also say that the stability is preserved while we add a
fourth laser beam because the phase and wave vector of this laser
beam are related to the phases and wave vectors of the other laser
beams by the same linear combination, as shown by
Eqs.~(\ref{equ:Lattice_phase_relation}) and
(\ref{equ:Lattice_kj_relation}). Indeed, if we consider the
configuration of Fig.~\ref{fig:Lattice_abc_lattice_schemes}(b),
the relation (\ref{equ:Lattice_kj_relation}) is still satisfied,
but the relation (\ref{equ:Lattice_phase_relation}) is not since
the phases are all independent. Therefore, we have
$\mathrm{rank}(\mathcal{K})=2<\mathrm{rank}\left(\mathcal{K}|\mathbf{\Phi}\right)=3$
and the optical lattice is not intrinsically phase stable.

\section{Discussion}

\subsection{Optical lattices in 1D and 3D}

Although all the examples given above were 2D lattices, everything
we have said is still true in other dimensions. For dimension
$d=1$, the minimal and folded optical lattices are degenerate and
correspond to the usual optical molasses. In this case, the
intrinsic phase stability is obvious, even without the above
matrix analysis, because any displacement of the retro-reflecting
mirror automatically shifts the phase of the standing wave by a
corresponding amount since the electric field has a node on the
mirror surface. For dimension $d=3$, the minimal and folded
optical lattices are generalizations of the two-dimensional case.
\begin{figure}[t!]
  \centering
  \includegraphics[width=9cm]{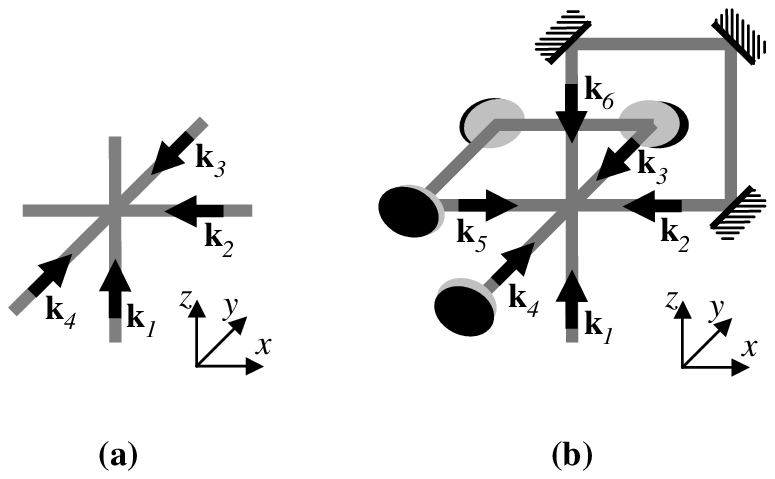}
  \caption{Three dimensional optical lattices:
    (a) minimal (with $d=l-1$);
    (b) folded.}
  \label{fig:3D_lattice_schemes}
\end{figure}
An example of a minimal optical lattice in three dimensions is
presented in Fig.~\ref{fig:3D_lattice_schemes}(a). This
configuration has been used by Treutlein {\em et al.} for
degenerate Raman sideband cooling \cite{Treutlein01}. Although the
geometry was not symmetrical, radiation pressure was reduced by
using a large detuning. Other examples of minimal optical lattices
are discussed in Ref.~\cite{Petsas94}. An example of a
three-dimensional folded optical lattice is presented in
Fig.~\ref{fig:3D_lattice_schemes}(b). This configuration is
obtained from configuration (a) by adding two laser beams, namely
$\mathbf{k}_5$ and $\mathbf{k}_6$. It is easy to show that the
phases and wave vectors of these two beams are related to the
other beams by the same linear relations:
\begin{eqnarray}
  \phi_5-\phi_1 &=& (\phi_4-\phi_1) + (\phi_3-\phi_1) -
  (\phi_2-\phi_1)\\
  \phi_6-\phi_1 &=& (\phi_4-\phi_1) + (\phi_3-\phi_1)
\end{eqnarray}
and
\begin{eqnarray}
  \mathbf{k}_5-\mathbf{k}_1 &=& (\mathbf{k}_4-\mathbf{k}_1) + (\mathbf{k}_3-\mathbf{k}_1) -
  (\mathbf{k}_2-\mathbf{k}_1)\\
  \mathbf{k}_6-\mathbf{k}_1 &=& (\mathbf{k}_4-\mathbf{k}_1) + (\mathbf{k}_3-\mathbf{k}_1)
  \;.
\end{eqnarray}
Therefore the optical lattice is intrinsically phase stable.

\subsection{Practical realisations of optical lattices}
\label{PracReal}

Work on cold atoms usually requires one to employ a symmetrical
beam configuration in order to avoid atoms being pushed aside by
radiation pressure. Minimal optical lattices can be designed in a
symmetrical geometry, but this requires a complex vacuum system.
For the 2D case, this means using 3 beams intersecting at
120$^{\circ}$ and a hexagonal coplanar geometry for the vacuum
system. To create a symmetrical 3D minimal lattice, the 4 beams
should form a regular tetrahedron. This adds even further to the
complexity of the vacuum apparatus (see Ref.~\cite{Shimizu91} for
an example of a tetrahedral magneto-optical trap).

Folded lattices, on the other hand, involve more beams but have a
more user-friendly geometry with beams intersecting at
right-angles. They can be aligned by auto-collimation and they
have the inherent advantage of balanced radiation pressure. In
addition, it is straightforward to adapt them to a power recycling
geometry, a tremendous advantage in many cases.

\subsection{Atomic beam collimation with a folded lattice}
\label{LaserCooling}

In a recent experiment, we have used the 2D folded optical lattice
of Fig.~\ref{fig:Lattice_abc_lattice_schemes}(a) to perform the
collimation of a continuous cesium beam in a fountain geometry
\cite{DiDomenico04}. In this folded lattice, we have realized
Zeeman-shift degenerate-Raman-sideband cooling in a continuous
mode. This powerful cooling technique allowed us to reduce the
atomic beam transverse temperature from 60~$\mu$K to 1.6~$\mu$K in
a few milliseconds. Remark that in this context, power recycling
is a big advantage since a high power is necessary to create a far
off-resonance optical lattice.

With the same experimental setup, we also realized collimation of
the continuous cesium beam using Sisyphus-like cooling in a 2D
optical lattice. We have experimented with the lattice
configurations (a) and (b) of
Fig.~\ref{fig:Lattice_abc_lattice_schemes} and the results are
summarized in table \ref{tab:temperatures}.
\begin{table}[tbp]
\caption{\label{tab:temperatures}Summary of transverse
temperatures obtained in the collimation experiment with
Sisyphus-like cooling in lattice configurations (a) and (b) of
Fig.~\ref{fig:Lattice_abc_lattice_schemes}. See subsection
\ref{LaserCooling} for details.}
\begin{center}
\begin{tabular}{lc}
  \hline\hline
  Optical lattice configuration  & Transverse temperature \\
                                 & ($\mu$K) \\
  \hline
  (a) Folded                     & 3.6(2)         \\
  (b) Unstable                   & 7.3(5)         \\
  \hline\hline
\end{tabular}
\end{center}
\end{table}
The best collimation has been obtained with the folded optical
lattice.

\subsection{Multi-color optical lattices}

Before concluding, we should like to point out that fulfilling the
condition (\ref{equ:rank_stability_condition}) is by no means the
only solution to get rid of the instability problem. Another
possibility is to average over the phase difference. To illustrate
this, consider the unstable 2D optical lattice of
Fig.~\ref{fig:Lattice_UnstableLattice}. If the two molasses have
different laser frequencies, the phase difference changes rapidly,
and the atoms see an optical lattice which is the average of the
optical shift over the phase variable. This solution has been used
with success by other groups \cite{Weiss00,Esslinger04}.

\section{Conclusion}

In this article we have established a necessary and sufficient
condition for the intrinsic phase stability of an optical lattice.
We have presented two practical solutions to fulfill this
condition, namely minimal and folded optical lattices. We have
shown that the minimal optical lattices, introduced for the first
time by Grynberg {\em et al.} in 1993, are sufficient but not
necessary for stability. Indeed, another possibility is to use a
folded optical lattice, as proposed by Rauschenbeutel {\em et al.}
in 1998. We have presented a particular example of folded optical
lattice, which has the advantages of power recycling, symmetry,
and a more convenient geometry. Henceforth, such a lattice would
seem to be a more natural choice for most experiments. Indeed, for
many applications a folded lattice looks like a better source of
cold atoms than a conventional six-beam optical molasses.

\section*{Acknowledgments}

This work was supported by the Swiss National Science Foundation,
the Swiss Federal Office of Metrology and Accreditation (METAS),
the canton of Neuch\^{a}tel, and the Swiss Confederation. AVT and VIYu
were partially supported by a grant INTAS-01-0855 and by RFBR
through grant \#04-02-16488.



\end{document}